\documentclass[12pt]{article}
\usepackage[dvips]{graphicx}
\usepackage{times}
\usepackage{epsfig}
\usepackage{amssymb}
\usepackage{amsmath}
\usepackage{multirow}

\begin{document}
\begin{center}
{\large\bf   
EIT phenomenon for the three-level hydrogen atoms and its application to the era of cosmological recombination\bigskip\\} 
${}^{a}$D. Solovyev and ${}^{b,c}$V. Dubrovich
\medskip\\    
$^a$ Department of Physics, St. Petersburg State University 198504, St. Petersburg, Russia\\
$^b$ St. Petersburg Branch of Special Astrophysical Observatory, Russian Academy of Sciences, 196140, St. Petersburg, Russia \\
$^c$ Nizhny Novgorod State Technical University n. a. R. E Alekseev, LCN, GSP-41, Minin str. 24, 603950, N. Novgorod, Russia\\
E-mail: solovyev.d@gmail.com
\end{center}

\begin{abstract}
The paper evaluates the contribution of the electromagnetically induced transparency (EIT) phenomenon to the processes of the microwave background (CMB) formation in early universe. We found the additional function $f$ to the integrated line absorption coefficient. This makes it the necessity to upgrade the Sobolev escape probability: $p_{ij}(\tau_S)\rightarrow p_{ij}\left(\tau_S\cdot\left(1+f\right)\right)$. We calculated the magnitude of the function $f$ for different schemes of the hydrogen atom in the three-level approximation in terms of the field parameters. The electric field amplitudes are defined using the CMB distribution. We found that the contribution of $f$ can be significant in some cases.
\end{abstract}

\section{Introduction}

Electromagnetically induced transparency (EIT) is a well-known quantum interference phenomenon that arises when coherent optical fields couple to the states of a quantum system \cite{Marangos}. Interference effects arise because in quantum mechanics probability amplitudes must be summed and squared to obtain the total transition probability between the relevant quantum states. The interference effects can lead to profound modification of the optical and nonlinear optical properties of a medium. In EIT, the interference occurs between alternative transition pathways driven by radiation fields. The most interesting example of this type of interference is the cancellation of absorption for a probe field, which leads to an almost transparent medium for the corresponding spectral line. In this paper we examine the transparency of the primordial medium in the context of cosmological recombination.

Initially the medium is not transparent for the resonant spectral lines. The resonant photon can be scattered many times before it escapes from the medium. The photon escape from its environment can be described in terms of the radiation transfer theory. The "standard" calculation of the radiation transfer in the early universe is limited to the evaluation of the three-level hydrogen atom which includes the ground state, the first excited state and the continuum, see \cite{Zeld}, \cite{Peebles}. However, in view of recent astrophysical observations of CMB anisotropy solving the problem of radiation transfer for this kind of the three-level atom would be a hopeless approximation. It was found \cite{Seager2} that in conjunction with the Sobolev approximation \cite{Sobolev} the problem of line transfer in the medium for the 300-level model of atom gives more adequate description of CMB. 

Within the framework of the Sobolev approximation the escape probability $p_{ij}$ is introduced. $p_{ij}$ represents the probability of photons associated with the transition $j\rightarrow i$, where $j$ is the upper level and $i$ is the lower level, escaping from the enviroment without being further scattered or absorbed. If $p_{ij}=1$, the photons produced in the line transition escape to infinity - they contribute no distortion to the radiation field. If $p_{ij}=0$, no photons escape to infinity; all of them get reabsorbed, and the line is optically thick. In general, $p_{ij}\ll 1$ for the Lyman lines and $p_{ij}=1$ for all other line transitions \cite{Seager2}. The Sobolev escape probability is defined by the optical depth $\tau_S$ which indicates the quantity of light removed from a beam by scattering or absorption as it passes through a medium, $p_{ij}=p_{ij}(\tau_S)$.

However, this kind of the approximation does not allow taking into account the external field influence on the properties of the medium. In terms of Sobolev approximation the optical depth is defined by the Einstein coefficients for the isolated atom: 
\begin{eqnarray}
\label{1a}
\tau_S = \frac{A_{ji}\lambda_{ij}^3[n_i(g_j/g_i)-n_j]}{8\pi H(z)}\, ,
\end{eqnarray}
where $H(z)$ is the Hubble expansion rate, $\lambda_{ij}$ is the central line wavelength, $n_i$ and $g_i$ are the actual populations and the degeneracy of $i$th states of the atom, respectively. 

In principle, Eq. (1) reflects the simplest case when, for example, the one-photon absorption process is described. However, the presence of external field can bring about effects of the next order (see below) since it changes the medium characteristics, the absorption in particular. In \cite{SDP} it was demonstrated that the function $f$ is added to the optical depth $\tau_S$ when the EIT phenomenon is taken into account. It was shown that the contribution to the photon escape probability can reach a level of $1.5\%$. 

Similarly to \cite{SDP}, but with a larger set of schemes, we use the three-level approximation for the description of the EIT phenomenon in the recombination era of the early universe. We consider the hydrogen atom having a set of ground and two excited states corresponding to the principal quantum number $n\leq 3$. All of the three-level systems discussed can be reduced to one of a number schemes depicted on Fig. 1.

\section{Theory and basic equations}

In \cite{Dubr}, the contribution of the decays of the atomic states with the principal number $n>2$ was discussed in context of the CMB formation processes. It was shown that two-photon transitions from the $nl$ excited states to the ground level in hydrogen atom can lead to a $1\%$-level distortion of the microwave background. Such modifications should bring a strong impact to the determination of the key cosmological parameters \cite{Lewis}. 

As a rule, the astrophysical investigations of the CMB in the spectral lines comes down to the calculations of the direct (emission) and inverse (absorption) processes in atom. The absorption and emission processes are related by the principle of local thermodynamic equilibrium (LTE). The LTE principle requires that the direct and inverse processes were exactly the same. This means that in the external Planck radiation field the photon, which is complementary to specified one, can always be found. Thus, absorption of a photon in one- or multiphoton process is regulated by the decay rate.

The other kind of multiphoton processes can be described within the framework of atom-field interaction when the external field acts on two adjacent resonances in atom. In particular, the mechanism of Velocity Selective Coherent Population Trapping (VSCPT) effect \cite{Asp} becomes possible. Velocity selection arises because the sensitivity to Doppler shift is weak (LTE principle), Doppler reduction of frequency for the one of photons is compensated by the corresponding increase of frequency for other one. Recently, it was shown also that the EIT phenomenon should be taken into account by means of upgrade of the Sobolev escape probability \cite{SDP}. In accordance to \cite{Seager2}, the Sobolev optical depth should be defined in terms of the Einstein coefficients. The equation (\ref{1a}) defines the absorption coefficient or the opacity $k$ with the separated normalized line profile $\phi(\nu_{ij})$, i.e. the optical depth is defined with the integrated line absorption coefficient $\tilde{k}$, $k=\tilde{k}\phi(\nu_{ij})$. Thus the Sobolev optical depth can be presented in the form:
\begin{eqnarray}
\label{2}
\tau_S=\frac{\lambda_{ij}\tilde{k}}{|\textit{v}'|}\, ,
\end{eqnarray}
where $\textit{v}'$ is the velocity gradient which is given by the Hubble expansion rate $H(z)$, see \cite{Seager2}.

On the other hand, the absorption coefficient depends strongly on the external conditions. In presence of an external field the opacity can be defined from the density matrix $\rho_{ij}$ as follows:
\begin{eqnarray}
\label{3}
k=\frac{N d_{ij}^2\omega_{ij}}{2\varepsilon_0\Omega_{ij}}{\rm Im}\big\{\rho_{ij}\big\}\, ,
\end{eqnarray}
where $\varepsilon_0$ is the vacuum permittivity, $N$ is the number of atoms, $d_{ij}$ is the dipole matrix element, $\Omega_{ij}$ is the Rabi frequency, $\omega_{ij}$ defines the frequency of the corresponding transition $|j>\rightarrow |i>$ and ${\rm Im}\big\{\rho_{ij}\big\}$ is the imaginary part of the density matrix.

In zeroth approximation, the imaginary part of density matrix is expressed through the Lorentz line profile that leads to the standard definition of the absorption coefficient \cite{Seager2}. However, Eq. (\ref{3}) allows taking into account the influence of the external field. This fact is determined by the presence of additional terms in the imaginary part of $\rho_{ij}$.

To define the field amplitudes we consider the next picture: after recombination, released photons were able to travel through the universe with the creation of the primordial background radiation. In present time, puzzling laboratory observations which have been attributed to photoionization, field-ionization anomalies, and collisional ionization can be easily explained by blackbody-radiation-induced transitions. The blackbody-induced decay, excitation and ionization rates in hydrogen and helium atoms were evaluated in \cite{GNO}. For the first time, the calculation and the experimental observation of the mentioned above phenomena were reported in \cite{GC}. Thus, this kind of influence of the CMB radiation on the corresponding narrowband transition line should be included in the description of cosmological recombination of the universe. The temperature of the recombination era is well known from laboratory experiments, $T_e\approx 4500-3000$ K. Therefore, the field amplitudes for a circular polarized wave can be obtained from the (thermal-averaged) spectral energy density
\begin{eqnarray}
\label{17}
\frac{c \varepsilon_0 |E|^2}{4\pi}=\frac{2h\nu_{ij}^3\Delta\nu_{ij}}{c^2}\frac{1}{e^{\frac{h\nu_{ij}}{k_B T_e}}-1}\, ,
\end{eqnarray}
where $c$ is the speed of light, $k_B$ is the Boltzmann constant, $h$ is Planck's constant and in further calculations we employ $T_e=3000$ K. The right-hand side of the equation above corresponds to the black-body distribution of the CMB, while left-hand side defines the (electrical) energy density. 

Due to the effect of dephasing, the question of the spectral line broadening arises with the change of the $\Delta \nu_ {ij}$ in Eq. (\ref{17}). To prevent dephasing phenomena we consider the external perturbation as a photon with the frequency varying in narrow strip. We should note that it is always possible in view of the continual character of the CMB distribution. Taking in mind the energy conservation law the width of the corresponding transition appears as the natural parameter. In our calculations we employ the following relation $\Delta\nu_{ij} = \Delta_{p}(\Delta_{c})\sim\Gamma_i$.

\section{Derivation of the imaginary part of the density matrix element}

To describe the system response on the external field we consider the three-level atom which is constructed from the ground and two excited states. Taking in consideration the continual character of the cosmic microwave background, we can always pick up the resonant frequency even if the detuning vanishes, and, thus, avoid the Doppler effect. Moreover, the description of the "atom-field" system can be reduced to the consideration of the spontaneous decays. The collisional excitation and the ionization processes can be omitted at the relevant temperatures and densities \cite{Seager2}. In this section we employ the density matrix theory in three-level approximation. The density matrix definition and its applications can be found, for example, in \cite{Weiner}, \cite{Boyd}.

The time evolution of the density matrix elements in the rotating wave approximation is given by the master equations (in atomic units):
\begin{eqnarray}
\label{13}
\dot{\rho}_{12}=-(\gamma_{12}-i(\Delta_p+\Delta_c))\rho_{12}-i\Omega_p\rho_{32}+i\Omega_c\rho_{13}\, ,
\nonumber
\\
\dot{\rho}_{23}=-(\gamma_{23}+i\Delta_c)\rho_{23}+i\Omega_c(\rho_{22}-\rho_{33})+i\Omega_p\rho_{21}\, ,
\\
\nonumber
\dot{\rho}_{13}=-(\gamma_{13}-i\Delta_p)\rho_{13}+i\Omega_p(\rho_{11}-\rho_{33})+i\Omega_c\rho_{12}\, ,
\end{eqnarray}
where the complex conjugate equations for the density matrix elements $\rho_{21}$, $\rho_{31}$ and $\rho_{32}$ are omitted for brevity. In Eqs. (\ref{13}) $\Omega_p$, $\Omega_c$ are the Rabi frequencies of the probe and coupling fields, respectively. The $\Omega=\bold{d}_{ij} \bold{E}$, where $\bold{E}$ is the electric field strength, the $\omega_p$, $\omega_c$ are the frequencies of the electric fields $\bold{E}$ for the corresponding transition lines $j\rightarrow i$. Widths $\gamma_{ij}$ can be defined by $\gamma_{ij}=\frac{1}{2}(\Gamma_i+\Gamma_j)$, where $\Gamma_i$ denotes the total decay width of the $i$th level. Finally, $\Delta_p=\omega_3-\omega_1-\omega_p$ and $\Delta_c=\omega_3-\omega_2-\omega_c$ are the corresponding detunings, where $\omega_i$ with $i=1,2,3$ are the energies of the atomic states. 

Equations (\ref{13}) can be solved in the adiabatic regime or the steady-state approximation, i.e. $\dot{\rho}_{ij}=0$ which assumes the utilization of weak fields (the populations of the states are changed weakly), \cite{Whitley}-\cite{Wiel}. To find the solution we consider the case of the full population of the ground state as the initial condition:
\begin{eqnarray}
\label{14}
\rho_{11}(0)=1
\\
\nonumber
\rho_{22}(0)=\rho_{33}(0)=0.
\end{eqnarray} 

For the $\Xi$ (cascade or ladder) scheme the matrix elements $\rho_{21}$, $\rho_{31}$ and $\rho_{32}$ can be found in the form:
\begin{eqnarray}
\label{15}
\rho_{31}=-\frac{i\Omega_p\left[(\gamma_{12}+i\delta)(\gamma_{23}+i\Delta_c)+\Omega_p^2\right]}{(\gamma_{13}+i\Delta_p)\left[(\gamma_{12}+i\delta)(\gamma_{23}+i\Delta_c)+\Omega_p^2\right]+(\gamma_{23}+i\Delta_c)\Omega_c^2}\, ,
\\
\nonumber
\rho_{32}=-\frac{i\Omega_p^2\Omega_c}{(\gamma_{13}+i\Delta_p)\Omega_p^2+(\gamma_{23}+i\Delta_c)\left[(\gamma_{12}+i\delta)(\gamma_{13}+i\Delta_p)+\Omega_c^2\right]}\, ,
\\
\nonumber
\rho_{21}=-\frac{(\gamma_{23}+i\Delta_c)\Omega_p\Omega_c}{(\gamma_{13}+i\Delta_p)\Omega_p^2+(\gamma_{23}+i\Delta_c)\left[(\gamma_{12}+i\delta)(\gamma_{13}+i\Delta_p)+\Omega_c^2\right]}\, ,
\end{eqnarray}
where $\delta=\Delta_p+\Delta_c$. It should be noted that Eqs. (\ref{13}) can be easily converted to the $\Lambda$ (Lambda) and $V$ (Vee) schemes by the replacement $\Delta_c\rightarrow -\Delta_c$ and $\Delta_p\rightarrow -\Delta_p$, respectively.

In \cite{Whitley}-\cite{Wiel} the expansion in a power of the $\Omega_p$ and $\Omega_c$ for the solution of Eqs. (\ref{13}) was demonstrated in case of the zeroth detunings (resonance approximation). The same procedure can be carried out in our case, but in this paper we keep the dependence on detunings. The reason is the extracting the line profile in the absorption coefficient, Eq. (\ref{3}). For the brevity of paper we present only the final result. Separating out the Lorentz profile, the imaginary part of the $\rho_{31}$ element is written as:
\begin{eqnarray}
\label{16}
{\rm Im}\{ \rho_{31} \}\approx \frac{\gamma_{13}\Omega_p}{\Delta_p^2+\gamma_{13}^2}\left[1+f(\Omega_p,\Omega_c,\Delta_p,\Delta_c)\right]\, ,
\end{eqnarray}
where function $f(\Omega_p,\Omega_c,\Delta_p,\Delta_c)$ in the first order of the $\Omega_p$, $\Omega_c$ is
\begin{eqnarray}
\label{16a}
f(\Omega_p,\Omega_c,\Delta_p,\Delta_c)\approx \frac{\gamma_{13}(\Delta_c^2-\gamma_{12}^2)+2\gamma_{12}\Delta_c\delta}{\gamma_{12}(\gamma_{12}^2+\Delta_c^2)(\gamma_{13}^2+\delta^2)}\Omega_c^2+...
\end{eqnarray}
In \cite{SDP}, the similar expression for the function $f$ can be found in case of the $\Xi$-scheme. The complete analytical representation of the function $f$ is too cumbrous and further we operate with the numerical values only. We omit also the accurate consideration of radiation transfer and adduce only the estimates in assumption of the parametrical dependence on the detunings. We should note, the matrix elements $\rho_{23}$ and $\rho_{21}$ make a much smaller contribution since they are suppressed by the factor $\Omega_p$.

\section{Numerical evaluation}

In this section we evaluate the function $f(\Omega_p,\Omega_c,\Delta_p,\Delta_c)$ numerically. The physical meaning of the $f$ can be found in \cite{Wiel} with the use of Fig. 2. The processes corresponding to the function $f$ are illustrated on the Fig. 2 b), c) and d). Namely, the a) graph corresponds to the "standard" Sobolev escape probability (see Eqs. (31), (33) in \cite{Seager2}). However, the coupling field leads to the delay of the electron on the excited states and impedes the final recombination in view of the additional processes, see graphs b)-d). Function $f$ is the dimensionless, it represents the contribution of the multiphoton processes b)-d) in respect to the one-photon absorption process a).

\subsection{Ladder or cascade scheme $\Xi$}

Recently the cascade $1s-2p-3s$ scheme for the hydrogen atom was considered \cite{SDP}. It was shown that in the case of exact resonances (when both detunings $\Delta_p$ and $\Delta_c$ are equal to zero) the maximal value of the $f$, Eqs. (\ref{16})-(\ref{16a}), is about $1.5\%$ and reaches $0.95\%$ for the two-photon resonance, when $\Delta_p+\Delta_c=0$ and $\Delta_p\neq0$, $\Delta_c\neq0$.

Similar to \cite{SDP}, we consider the cascade scheme but with the taking into account of the fine structure of the hydrogen atom. Taking into consideration the Lamb shift, the three-level ladder scheme can be defined as follows: $|1>\equiv|1s>$, $|3>\equiv |2p_{1/2}>$ and $|2>\equiv|2s>$. Substituting the corresponding widths and frequencies $\Gamma_{2s}=8.229$ $s^{-1}$, $\Gamma_{2p}=6.26826\cdot 10^8$ $s^{-1}$, and $\nu_{31}=2.4674\cdot 10^{15}$ $Hz$, $\nu_{23}=1.057911\cdot 10^9$ $Hz$ in Eq. (\ref{17}) and Eq. (\ref{16}) with $\Delta\nu_{ij}= \Gamma_{2p}$, we obtain
\begin{eqnarray}
\label{18}
E_{p}\approx 0.000068802\,\, V/m = 1.33799\cdot 10^{-16}\, a.u.,
\\
\nonumber
E_{c}\approx 0.0017496\,\, V/m = 3.40242 \cdot 10^{-15}\, a.u.,
\\
\nonumber
\Omega_p\approx 9.96713\cdot 10^{-17}\, a.u.,
\\
\nonumber
\Omega_c\approx 1.02073\cdot 10^{-14}\, a.u.
\end{eqnarray}
The numerical results of the function $f$ for the different magnitudes of detunings are compiled in Table 1.
\begin{table}[h!]
{\citation\, Table 1. The numerical values of the function $f(\Omega_{p}, \Omega_{c}, \Delta_{p}, \Delta_{c})$ for the different magnitudes of detunings are presented. In the first column the values of the $f(\Omega_{p}, \Omega_{c}, \Delta_{p}, \Delta_{c})$ are listed, in the second and third columns the detunings are given.}
\begin{center}
\begin{tabular}{c|c|c}
\hline
\hline
$f\left(\Omega_{p},\Omega_{c},\Delta_{p},\Delta_{c}\right)$ & $\Delta_{p} \,\,s^{-1}$  & $\Delta_{c}\,\,s^{-1}$  \\
 \hline
$-1.38066\cdot 10^{-4}$ & $0$ & $0$ \\
 \hline
$-1.38066\cdot 10^{-4}$ & $\Gamma_{2s}=8.229$ & $-\Gamma_{2s}=-8.229$ \\
\hline
$8.28426\cdot 10^{-5}$ & $\Gamma_{2p}=6.26826 \cdot 10^8$ & $-\Gamma_{2p}=-6.26826 \cdot 10^8$ \\
 \hline
$-8.12364\cdot 10^{-6}$ & $\Gamma_{2s}=8.229$ & $\Gamma_{2s}=8.229$ \\
 \hline
$7.25115 \cdot 10^{-13}$ & $\Gamma_{2p}=6.26826 \cdot 10^8$ & $\Gamma_{2s}=8.229$ \\

 \hline
$3.62557\cdot 10^{-13}$ & $\Gamma_{2p}=6.26826 \cdot 10^8$ & $\Gamma_{2p}=6.26826 \cdot 10^8$ \\
  \hline
  \hline
\end{tabular}
\end{center}
\end{table}

In particular, in the case of exact resonances the maximal value of the $f$ is $-1.38066\cdot 10^{-4}$, i.e. $0.01\%$. In the case of exact two-photon resonance $\Delta_p+\Delta_c=0$, when the frequencies of fields are close but differ slightly from the corresponding resonances, $f\approx 0.0083\%$. Thus, in contrast to results of \cite{SDP}, the maximal value of the function $f$ is beyond the required accuracy of the CMB calculations. We should note also if $\Delta\nu_{ij}= \Gamma_{2s}$ in Eq. (\ref{17}) then the maximal value of $f$ is of the order of $10^{-12}$. We can conclude, the EIT phenomenon does not require the consideration of the fine structure for the $\Xi$ scheme.

\subsection{Lambda scheme $\Lambda$}

To describe the $\Lambda$ scheme, Fig. 1b), the substitution $\Delta_c\rightarrow -\Delta_c$ should be performed in Eqs. (\ref{13}). Using the set of states $|1> \equiv |1s>$, $|2> \equiv |2s>$, $|3> \equiv |3p>$ in conjunction with $\nu_{21}=2.4674\cdot 10^{15}$ $Hz$, $\nu_{31}=2.9243\cdot 10^{15}$ $Hz$, $\Gamma_2\equiv\Gamma_{2s}=8.229$ $s^{-1}$ and $\Gamma_3\equiv\Gamma_{3p}=1.89802 \cdot 10^8$ $s^{-1}$, assuming also $\Delta \nu_{32}\approx \Gamma_{2s}$ and $\Delta \nu_{31}\approx\Gamma_{3p}$ in Eq. (\ref{17}), we obtain
\begin{eqnarray}
\label{19}
E_{p}\approx  2.45765 \cdot 10^{-18}\, a.u.,
\\
\nonumber
E_{c}\approx  1.17805 \cdot 10^{-14}\, a.u.,
\\
\nonumber
\Omega_p\approx 7.33142\cdot 10^{-19}\, a.u.,
\\
\nonumber
\Omega_c\approx 2.08452 \cdot 10^{-14}\, a.u.
\end{eqnarray}
The numerical values of the function $f\left(\Omega_{p},\Omega_{c},\Delta_{p},\Delta_{c}\right)$ are given in Table 2.
\begin{table}[h!]
{\citation\, Table 2. The numerical values of the function $f(\Omega_{p}, \Omega_{c}, \Delta_{p}, \Delta_{c})$ for the different magnitudes of detunings are presented. In the first column the values of the $f(\Omega_{p}, \Omega_{c}, \Delta_{p}, \Delta_{c})$ are listed, in the second and third columns the detunings are given.}
\begin{center}
\begin{tabular}{c|c|c}
\hline
\hline
$f\left(\Omega_{p},\Omega_{c},\Delta_{p},\Delta_{c}\right)$ & $\Delta_{p} \,\,s^{-1}$  & $\Delta_{c}\,\,s^{-1}$  \\
 \hline
$-0.00189829$ & $0$ & $0$ \\
 \hline
$-0.00189829$ & $\Gamma_{2s}=8.229$ & $\Gamma_{2s}=8.229$ \\
\hline
$0.00113955$ & $\Gamma_{3p}=1.89802 \cdot 10^8$ & $\Gamma_{3p}=1.89802 \cdot 10^8$ \\
 \hline
$-0.000112064$ & $\Gamma_{2s}=8.229$ & $-\Gamma_{2s}=-8.229$ \\
  \hline
$3.29832\cdot 10^{-11}$ & $\Gamma_{3p}=1.89802 \cdot 10^8$ & $\Gamma_{2s}=8.229$ \\
 \hline
$1.64916\cdot 10^{-11}$ & $\Gamma_{3p}=1.89802 \cdot 10^8$ & $-\Gamma_{3p}=-1.89802 \cdot 10^8$ \\
  \hline
  \hline
\end{tabular}
\end{center}
\end{table}

If we assume $\Delta \nu_{32}\approx \Gamma_{2s}$ and $\Delta \nu_{31}\approx\Gamma_{2s}$ in Eq. (\ref{17}) then $\Omega_p\approx 1.52655 \cdot 10^{-22}$ and  $\Omega_c\approx 2.08452 \cdot 10^{-14}$ (in atomic units) can be obtained. The magnitude $\Omega_c$ is the dominant and coincides with the value in Eqs. (\ref{19}). Thus the same results as in Table 2 can be reproduced.

In case of $\Delta \nu_{32}\approx \Gamma_{3p}$ and $\Delta \nu_{31}\approx\Gamma_{3p}$, we receive
\begin{eqnarray}
\label{20}
\Omega_p\approx 7.33142 \cdot 10^{-19}\, a.u.,
\\
\nonumber
\Omega_c\approx 1.00111 \cdot 10^{-10}\, a.u.
\end{eqnarray}

These estimates show that we can neglect by terms of higher order in $\Omega_p$. However, the approximation Eq. (\ref{16}) is satisfied at the condition $\Omega_i/\gamma_{ij}\ll 1$ which is violated for the $\Omega_c/\gamma_{12}$. Then the solution of Eqs. (\ref{15}) should be used without series expansion. For the matrix element $\rho_{31}$ we can write
\begin{eqnarray}
\label{21}
\rho_{31}\approx -\frac{i\Omega_p}{\gamma_{13}+i\Delta_p+\frac{\Omega_c^2}{\gamma_{12}+i\delta}}.
\end{eqnarray}
The values of the function $f$ are listed in Table 3.

\begin{table}[h!]
{\citation\, Table 3. The notations are the same as in Tables 1.}
\begin{center}
\begin{tabular}{c|c|c}
\hline
\hline
$f\left(\Omega_{p},\Omega_{c},\Delta_{p},\Delta_{c}\right)$ & $\Delta_{p} \,\,s^{-1}$  & $\Delta_{c}\,\,s^{-1}$  \\
 \hline
$0.000761156$ & $\Gamma_{3p}=1.89802 \cdot 10^8$ & $\Gamma_{2s}=8.229$ \\
\hline
$0.000380479$ & $\Gamma_{3p}=1.89802 \cdot 10^8$ & $-\Gamma_{3p}=-1.89802 \cdot 10^8$ \\
\hline
$-9.04486\cdot 10^{-7}$ & $\Gamma_{2s}=8.229$ & $\Gamma_{3p}=1.89802 \cdot 10^8$ \\
  \hline
  \hline
\end{tabular}
\end{center}
\end{table}

In accordance to the results of Tables 2 and 3, the maximal value of the $f$ can be found on the level of $0.2\%$. The contribution of this order is comparable to modern accuracy of the CMB observations.

Taking into account the fine structure of levels for the $\Lambda$ scheme, we can set $|1>\equiv|1s>$, $|2>\equiv|2s>$, $|3>\equiv|2p_{3/2}>$ and, therefore, $\nu_{31}\approx 2.46741\cdot 10^{15}$ $Hz$, $\nu_{32}\approx 9.96903\cdot 10^9$ $Hz$, $\Gamma_3\equiv\Gamma_{2p}=6.26826\cdot 10^8$ $s^{-1}$ and $\Gamma_{2}\equiv\Gamma_{2s}=8.229$ $s^{-1}$. To estimate the field amplitudes we employ $\Delta\nu_{31}\sim \Gamma_{2p}$, $\Delta\nu_{32}\sim\Gamma_{2s}$. Then
\begin{eqnarray}
\label{22}
\Omega_p\approx 9.96632 \cdot 10^{-17}\, a.u.,
\\
\nonumber
\Omega_c\approx 1.10204 \cdot 10^{-10}\, a.u.
\end{eqnarray}
The corresponding numerical values of the $f$ are listed in Table 4. Then we can conclude that the influence of the external field on the optical depth is negligible in this case. The same result can be received when $\Delta\nu_{31}\sim \Gamma_{2s}$ and $\Delta\nu_{32}\sim\Gamma_{2s}$.
\begin{table}[h!]
{\citation\, Table 4. The notations are the same as in Tables 1.}
\begin{center}
\begin{tabular}{c|c|c}
\hline
\hline
$f\left(\Omega_{p},\Omega_{c},\Delta_{p},\Delta_{c}\right)$ & $\Delta_{p} \,\,s^{-1}$  & $\Delta_{c}\,\,s^{-1}$  \\
 \hline
$-1.60962\cdot 10^{-10}$ & $0$ & $0$ \\
\hline
$-1.60962\cdot 10^{-10}$ & $\Gamma_{2s}=8.229$ & $\Gamma_{2s}=8.229$ \\
 \hline
$9.65771\cdot 10^{-11}$ & $\Gamma_{2p}=6.26826 \cdot 10^8$ & $\Gamma_{2p}=6.26826 \cdot 10^8$ \\
\hline
$-9.46834\cdot 10^{-12}$ & $\Gamma_{2s}=8.229$ & $-\Gamma_{2s}=-8.229$ \\
 \hline
$8.45246\cdot 10^{-19}$ & $\Gamma_{2p}=6.26826 \cdot 10^8$ & $\Gamma_{2s}=8.229$ \\
  \hline
  \hline
\end{tabular}
\end{center}
\end{table}

However, defining $\Delta\nu_{31}\sim \Gamma_{2p}$ and $\Delta\nu_{32}\sim\Gamma_{2p}$ in Eq. (\ref{17}), we get
\begin{eqnarray}
\label{23}
\Omega_p\approx 9.96632 \cdot 10^{-17}\, a.u.,
\\
\nonumber
\Omega_c\approx 9.61828 \cdot 10^{-14}\, a.u.
\end{eqnarray}
with the numerical results listed in Table 5.
\begin{table}[h!]
{\citation\, Table 5. The notations are the same as in Tables 1.}
\begin{center}
\begin{tabular}{c|c|c}
\hline
\hline
$f\left(\Omega_{p},\Omega_{c},\Delta_{p},\Delta_{c}\right)$ & $\Delta_{p} \,\,s^{-1}$  & $\Delta_{c}\,\,s^{-1}$  \\
 \hline
$-0.0121124$ & $0$ & $0$ \\
\hline
$-0.0121124$ & $\Gamma_{2s}=8.229$ & $\Gamma_{2s}=8.229$ \\
 \hline
$0.00729051$ & $\Gamma_{2p}=6.26826 \cdot 10^8$ & $\Gamma_{2p}=6.26826 \cdot 10^8$ \\
\hline
$-0.000729015$ & $\Gamma_{2s}=8.229$ & $-\Gamma_{2s}=-8.229$ \\
 \hline
$6.43847\cdot 10^{-11}$ & $\Gamma_{2p}=6.26826 \cdot 10^8$ & $\Gamma_{2s}=8.229$ \\
 \hline
$3.21924\cdot 10^{-11}$ & $\Gamma_{2p}=6.26826 \cdot 10^8$ & $-\Gamma_{2p}=-6.26826 \cdot 10^8$ \\
  \hline
  \hline
\end{tabular}
\end{center}
\end{table}

As it follows from Table 5 the maximal value of the function $f\left(\Omega_{p},\Omega_{c},\Delta_{p},\Delta_{c}\right)$ reaches the level of $1\%$. The contribution of this order is sufficiently high and, therefore, should be taken into account for the theoretical description of the cosmic microwave background.

\subsection{Vee scheme $V$}

In this section we consider vee $V$ scheme depicted in graph Fig. 1c). With the set of states $|1>\equiv |3p>$, $|2>\equiv |2p>$ and $|3>\equiv|1s>$, we have $\nu_{23}=2.4674\cdot 10^{15}$ $Hz$, $\nu_{13}=2.9243\cdot 10^{15}$ $Hz$, $\Gamma_2\equiv\Gamma_{2p}=6.26826\cdot 10^8$ $s^{-1}$ and $\Gamma_1\equiv\Gamma_{3p}=1.89802 \cdot 10^8$ $s^{-1}$. Assuming $\Delta \nu_{23}\approx \Gamma_{2p}$ and $\Delta \nu_{13}\approx\Gamma_{3p}$ in Eq. (\ref{17}), we obtain
\begin{eqnarray}
\label{24}
\Omega_p\approx 7.33142 \cdot 10^{-19}\, a.u.,
\\
\nonumber
\Omega_c\approx 9.96713 \cdot 10^{-17}\, a.u.,
\end{eqnarray}
The numerical values of the function $f\left(\Omega_{p},\Omega_{c},\Delta_{p},\Delta_{c}\right)$ are given in Table 6.
\begin{table}[h!]
{\citation\, Table 6. The notations are the same as in Tables 1.}
\begin{center}
\begin{tabular}{c|c|c}
\hline
\hline
$f\left(\Omega_{p},\Omega_{c},\Delta_{p},\Delta_{c}\right)$ & $\Delta_{p} \,\,s^{-1}$  & $\Delta_{c}\,\,s^{-1}$  \\
 \hline
$-5.70837 \cdot 10^{-16}$ & $0$ & $0$ \\
\hline
$4.8248\cdot 10^{-16}$ & $\Gamma_{3p}=1.89802 \cdot 10^8$ & $\Gamma_{3p}=1.89802 \cdot 10^8$ \\
\hline
$4.52944 \cdot 10^{-16}$ & $\Gamma_{3p}=1.89802 \cdot 10^8$ & $-\Gamma_{3p}=-1.89802 \cdot 10^8$ \\
\hline
$4.14603\cdot 10^{-16}$ & $\Gamma_{2p}=6.26826\cdot 10^8$ & $\Gamma_{2p}=6.26826\cdot 10^8$ \\
\hline
$-1.14167 \cdot 10^{-16}$ & $\Gamma_{3p}=1.89802 \cdot 10^8$ & $\Gamma_{2p}=6.26826\cdot 10^8$ \\
\hline
$1.76641 \cdot 10^{-17}$ & $\Gamma_{2p}=6.26826\cdot 10^8$ & $-\Gamma_{2p}=-6.26826\cdot 10^8$ \\
  \hline
  \hline
\end{tabular}
\end{center}
\end{table}

Finally, taking into consideration the fine structure, assuming  $|1>\equiv |2p_{3/2}>$, $|2>\equiv |2p_{1/2}>$ and $|3>\equiv|1s>$, we have $\nu_{23}=2.4674\cdot 10^{15}$ $Hz$, $\nu_{13}=2.46741 \cdot 10^{15}$ $Hz$, $\Gamma_2\equiv\Gamma_{2p_{1/2}}=6.26826\cdot 10^8$ $s^{-1}$ and $\Gamma_1\equiv\Gamma_{2p_{3/2}}=6.26826 \cdot 10^8$ $s^{-1}$. As in the case of estimates (\ref{24}) the corresponding numerical values of the $f$ are completely negligible. In case of $\Delta \nu_{13}\approx \Gamma_{2p}$ and $\Delta \nu_{23}\approx\Gamma_{2p}$ in Eq. (\ref{17}), we obtain the vanishing values again, $f\sim 10^{-16}$. Thus, in context of CMB formation processes, the electromagnetically induced transparency phenomenon is not important for the $V$ scheme.

\subsection{Full settling in the $2s$ state}

The $2s \leftrightarrow 1s$ two-photon decay process is able to substantially control the dynamics of cosmological hydrogen recombination \cite{Zeld}, \cite{Peebles}, allowing about $57\%$ of all the hydrogen atoms in the Universe to recombine through this channel. Taking into account the full settling of the $2s$ state in hydrogen, we demonstrate briefly the influence of the EIT phenomenon on the CMB formation processes. Then the conditions Eqs. (\ref{14}) should be replaced by $\rho_{2s}(0)=1$, but others are equal to zero. For the cascade scheme, Fig. 1 a), with the estimates  Eqs. (\ref{18}) we found that the maximum value of the $f$ is about $0.02\%$, see Table 7.
\begin{table}[h!]
{\citation\, Table 7. The notations are the same as in Tables 1.}
\begin{center}
\begin{tabular}{c|c|c}
\hline
\hline
$f\left(\Omega_{p},\Omega_{c},\Delta_{p},\Delta_{c}\right)$ & $\Delta_{p} \,\,s^{-1}$  & $\Delta_{c}\,\,s^{-1}$  \\
 \hline
$-0.000276132$ & $0$ & $0$ \\
\hline
$-0.000276132$ & $\Gamma_{2s}=8.229$ & $\Gamma_{2s}=-8.229$ \\
\hline
$-0.0000552386$ & $\Gamma_{2p}=6.26826 \cdot 10^8$ & $-\Gamma_{2p}=-6.26826 \cdot 10^8$ \\
\hline
$0.00003314$ & $-\Gamma_{2p}=-6.26826\cdot 10^8$ & $\Gamma_{2p}-\Gamma_{2s}=6.26826\cdot 10^8$ \\
\hline
$-0.0000162473$ & $\Gamma_{2s}=8.229$ & $\Gamma_{2s}=8.229$ \\
\hline
$7.25115\cdot 10^{-13}$ & $\Gamma_{2p}=6.26826 \cdot 10^8$ & $\Gamma_{2p}=6.26826\cdot 10^8$ \\
  \hline
  \hline
\end{tabular}
\end{center}
\end{table}

For the $\Lambda$ scheme we found that the maximal value of the $f\left(\Omega_{p},\Omega_{c},\Delta_{p},\Delta_{c}\right)$ for the matrix element $\rho_{32}$ and the estimates Eqs. (\ref{19}) is of the order $10^{-12}$. Taking into account the fine structure and values Eqs. (\ref{23}), the maximal value of the function $f$ on the level $10^{-8}$ can be found. We should stress that in the case of full population of the $2s$ state the matrix element $\rho_{31}$ is suppressed by the factor $\Omega_p\Omega_c^2$ and, therefore, the effect remains insignificant for the Ly$_\alpha$ transition in the $\Lambda$ scheme.

\section{Conclusions}

In the last decade the research of the CMB anisotropy was carried out by many authors, see, for example, \cite{Chluba}. In principle, the full set of states in hydrogen atom is required for the detailed description of the CMB anisotropy. Recently, to achieve an accuracy $0.1\%$ the set of states $n\geqslant 100$ of the hydrogen atom levels with the separate angular momentum substates was used in order to obtain sufficiently accurate recombination histories \cite{HH}. Investigations of the CMB formation is based on radiation transfer theory \cite{Seager2} and includes the Sobolev approximation.

However, within the framework of the radiation transfer theory, the influence of an external field on the characteristics of the environment has not yet been taken into account. We have considered the electromagnetically induced transparency phenomenon as a one of such occurrence. We found out that EIT phenomenon becomes apparent in terms of CMB formation and can reach the level of $1\%$. 

To describe the effect of electromagnetically induced transparency we have employed the quantum optical derivation of the integrated line absorption coefficient. In this case the absorption coefficient contains the additional terms. These terms depend on the external field. The magnitudes of fields were deduced from the CMB distribution. We have evaluated the contribution of the EIT phenomenon for the different schemes of the hydrogen atom in a three-level approximation. It is found that the additional function $f\left(\Omega_{p},\Omega_{c},\Delta_{p},\Delta_{c}\right)$ should be included into the Sobolev escape probability, i.e. $p_{ij}(\tau_S)\rightarrow p_{ij}\left(\tau_S\cdot[1+f\left(\Omega_{p},\Omega_{c},\Delta_{p},\Delta_{c}\right)]\right)$. 

In \cite{SDP} the maximum value of the $f$ function on the level of $1.5\%$ was found in case of exact resonances for the cascade scheme, Fig. 1 a). In this paper the maximal contribution of the EIT phenomenon was received for the $\Lambda$ scheme with the fine structure of levels and the full population of the ground state in hydrogen, see Table 5. For this case the maximal contribution of the $f$ is about $1.2\%$. 

Another significant contribution arises for the $\Lambda$ scheme when $|1>=|1s>$, $|2>=|2s>$, $|3>=|3p>$ and the estimates Eqs. (\ref{19}) are applied. The order of the $f\left(\Omega_{p},\Omega_{c},\Delta_{p},\Delta_{c}\right)$ is about $0.2\%$ in this case, see Table 2. For the vanishing detunings and estimates Eqs. (\ref{20}), we found the contribution of the order of $0.1\%$, see Table 3. In all other cases, the negligible values were obtained. 

Finally, in the case of the full settling of the $2s$ state we found that the EIT phenomenon does not contribute significantly. The maximal value of the $f$ arises for the cascade scheme and is about $0.03\%$. However, we should note the exponential behavior of the field amplitudes as a function of temperature. With the increasing of temperature $T_e$ the larger values of amplitudes can be obtained, see Eq. (\ref{17}). Therefore, the contribution of the EIT phenomenon should be more considerable for the higher temperatures $T_e$.

\begin{center}
Acknowledgments
\end{center}
The work was supported by RFBR (grant No. 08-02-00026,  No. 11-02-00168-a and No. 12-02-31010) and by The Ministry of education and science of Russian Federation, project 8420.

\begin{figure}[h]
\begin{center}
\includegraphics[width=14cm]{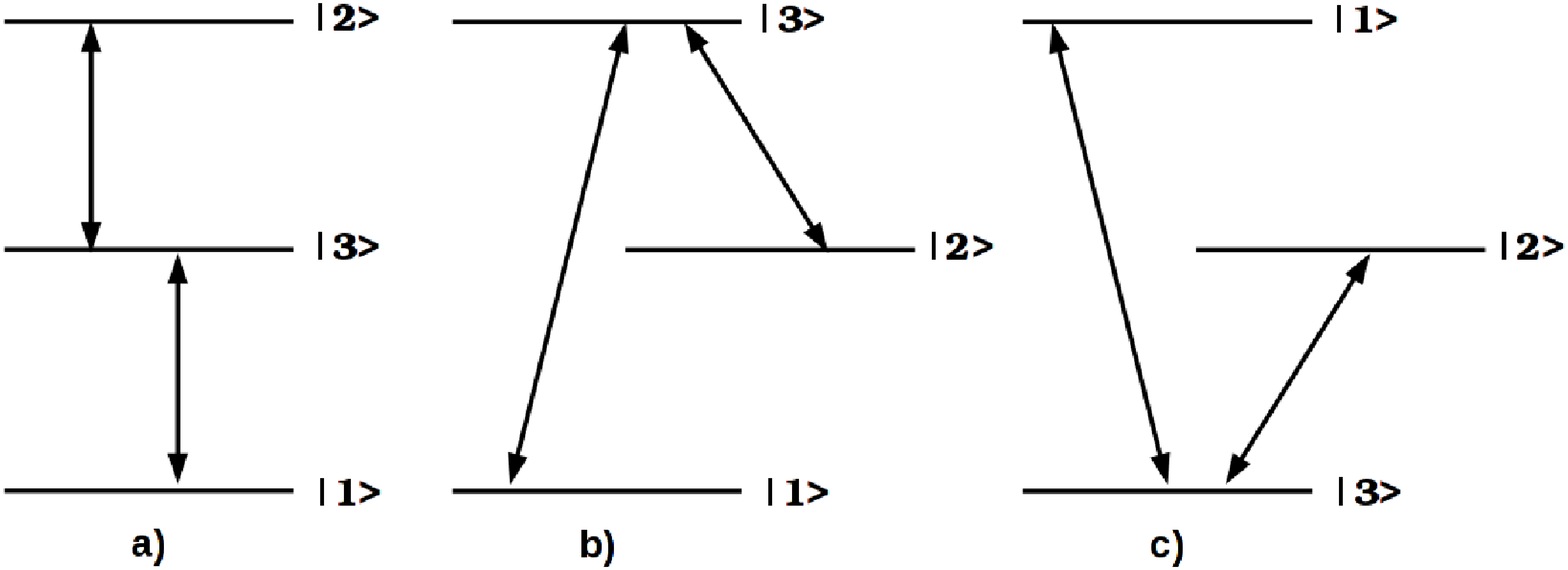}
\end{center}
\caption{\label{Fig1} Basic three-level schemes: a) ladder or cascade ($\Xi$) scheme with $E_1 < E_3 < E_2$, b) lambda ($\Lambda$) scheme with $E_1 < E_2 < E_3$, c) vee ($V$) scheme with $E_3 < E_1$ and $E_3 < E_2$.}
\end{figure}
\begin{figure}[h]
\begin{center}
\includegraphics[width=14cm]{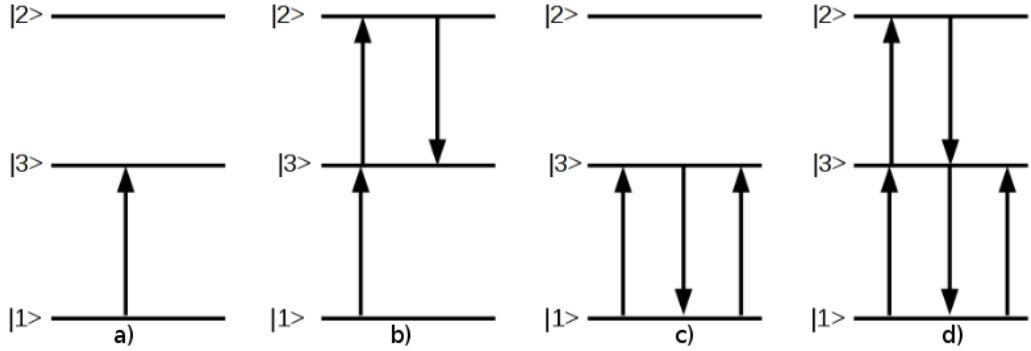}
\end{center}
\caption{\label{Fig2} The transition processes, occurring in the three-level ladder system and corresponding to the different term of the imaginary part of Eqs. (\ref{16}), are depicted: part a) of the figure represents the one-photon absorption processes which corresponds to the "standard" definition of the optical depth; part b), c) and d) represent the terms of the next orders in $\Omega_p$ and $\Omega_c$ appearing due to the presence of the external fields and derived via series expansion of the matrix elements Eqs. (\ref{15}). The function $f(\Omega_p,\Omega_c,\Delta_p,\Delta_c)$ Eq. (\ref{16a}) is presented by the graphs b), c) and d) \cite{Wiel}.}
\end{figure}
\end{document}